\documentclass{icrc}

\usepackage{times}
\usepackage{graphicx} 
\begin{document}

\title{ Analytical description of muon distributions at large depths}
\author[1]{S. Klimushin}
\author[1]{E. Bugaev}
\affil[1]{
Institute for Nuclear Research, Russian Academy of Science, Moscow 117312, Russia}
\author[1,2]{I. Sokalski}
\affil[2]{now at DAPNIA/SPP, CEA/Saclay, 91191 Gif-sur-Yvette CEDEX, France}
\correspondence{klim@pcbai11.inr.ruhep.ru}

\firstpage{1}
\pubyear{2001}


\maketitle

\begin{abstract}
The analytical expression for integral energy spectra and zenith angle
distributions of atmospheric muons at large depths is derived.
Fluctuations of muon energy losses are described using the 
parametrized correction factor. The fitting formula for the
sea level muon spectrum at different zenith angles for spherical
atmosphere is proposed. The concrete calculations for pure water are presented. 
\end{abstract}

\section{Introduction}

The last several years have been marked by the start of full scale data taking of large neutrino
and muon telescopes located at lake Baikal (NT-200) and in deep polar ice (AMANDA). 
Two underwater telescopes
ANTARES and NESTOR assuming installation at greater depths are being intensively 
constructed in the Mediterranean. The possibility of deployment of telescopes with huge detecting 
volumes up to 1 km$^3$ is also under wide investigation.

So, the knowledge of expected angular distribution of integral flux of atmospheric muons deep underwater  
is of interest not only for cosmic ray physics 
but also for the estimation of the possible background for neutrino detection and at last for a test 
of the correctness of underwater telescope data interpretation by using the natural flux of atmospheric muons as calibration source.    
The last item frequently implies the estimation with an appropriate accuracy (e.g., better than 5$\,\%$
for a given sea level spectrum) the 
underwater integral muon flux for various sets of depths, cut-off energies and angular bins 
especially for telescopes of big spacial dimensions.

Up to now the presentation of the results of calculations of muon propagation through thick layers of water 
both for parent muon sea level spectra (especially for angular
dependence taking into account the sphericity of atmosphere) and for underwater angular flux has not been 
quite convenient when applied to concrete underwater arrays.
In addition, a part of numerical results is available only in data tables (often insufficient for accurate interpolation) 
and figures.
The possibility of direct implementation of Monte Carlo methods depends on the availability of corresponding codes
and usually assumes rather long computations and accurate choice of the grid for simulation parameters to avoid big 
systematic errors.   
Therefore, there remains the necessity of analytical expressions for underwater muon integral flux. 
In addition, the possibility of reconstructing the parameters of a sea level
spectrum by fitting measured underwater flux in the case of their direct relation looks rather attractive. 
  
In this paper we present rather simple method allowing one to analytically calculate 
the angular distribution of integral muon flux
deep under water for cut off energies (1--$10^4)\,$GeV and slant depths of (1--16)$\,$km 
for conventional ($\pi,\,K$) sea level atmospheric muon spectra fitted by means 
of five parameters.  The fluctuations of muon energy losses are taken into account. 




\section{Basic formulas}
According to the approach of work~\citep{KBS} the analytical expression
for calculations of underwater angular integral flux above cut-off energy $E_f$ for a slant depth $R=h/ \cos \theta$ seen at vertical depth $h$ 
at zenith angle $\theta$ and allowing for the fluctuations of energy loss is based on the relation
\begin{equation}\label{ad1}
F_{fl}(\geq E_f,R,\theta)=\frac {F_{cl}(\geq E_f,R,\theta)} {C_{f}(\geq E_f,R,\theta)},  
\end{equation}
where correction factor $C_f$ is expressed, by definition, by the ratio of theoretical integral flux calculated in
the continuous loss approximation to that calculated by exact Monte Carlo. The analytical parametrization for $C_f$
is presented in Refs.~\citep{KBS,KBS1}. The dependencies of the correction factor on $E_f$ and $R$, calculated for any
reasonable sea level spectrum represent the set of rather smooth curves. 

The angular flux $F_{cl}(\geq E_f,R,\theta)$ based on effective 
linear continuous energy losses $\alpha + \beta E$ having 2 slopes, is calculated by the following rule: 
\begin{eqnarray} \label{ad2}
F_{cl}(\geq E_f,R,\theta) = \nonumber \\ 
\left\{
\begin{array}{ll} F_{cl}(\geq E_f,R,\theta;\alpha_{1},\beta_{1})
                         & \mbox{~for~}  R \leq R_{12}, \\
                F_{cl}(\geq E_{12},(R-R_{12}),\theta;~\alpha_{2},\beta_{2})
                         & \mbox{~for~}  R > R_{12}. 
\end{array}\right.
\end{eqnarray}
Here $E_{12}$ is the energy in the point of slope change from $(\alpha_1,\beta_1)$ to $(\alpha_2,\beta_2)$ and 
$R_{12}$ is the muon path from the energy $E_{12}$ till $E_f$ which is given by
$$R_{12}=\frac {1}{\beta_1}\,\ln \biggl( \frac {\alpha_{1}+E_{12}\beta_{1}}{\alpha_{1}+E_{f}\beta_1} \biggr).$$ 

The formula for integral muon angular flux in the assumption of linear 
continuous energy losses is as follows:
\begin{eqnarray}\label{ad3}
F_{cl}(\geq E_f,R,\theta;~\alpha,\beta)
=\frac{e^{-\beta R\gamma}}{\gamma} \nonumber \\ 
\times \sum_{i={\pi,K}} D_{0_i} E^{cr}_{0_i}(\theta) 
(E_f+y_i)^{-\gamma} (1-z_i)^{1-\gamma}\,\mbox{S}(z_{i},\gamma), 
\end{eqnarray}
where subscript $i$ stands over both pion ($\pi$) and kaon ($K$) terms and 
\begin{eqnarray*}
y_i = \frac{\alpha}{\beta}\,(1-e^{-\beta R}) + E^{cr}_{0_i}(\theta)\,e^{-\beta R}, \\
z_i = \frac {E^{cr}_{0_i}(\theta)\,e^{-\beta R}}{ E_{f}+y_i },
\qquad E^{cr}_{0_i}(\theta)=\frac {E^{cr}_{0_i}(0^\circ)}{\cos\theta^*},\\
\mbox{S}(z,\gamma) = 1+ \sum_{n=1}^{\infty} n!\,z^{n} \biggl (\prod_{j=1}^{n} (\gamma+j) \biggr )^{-1}= 
1+\frac{z}{\gamma+1}\\
+\frac{2z^2}{(\gamma+1)(\gamma+2)}+ 
\frac{6z^3}{(\gamma+1)(\gamma+2)(\gamma+3)} + \dots~. 
\end{eqnarray*}

When using expression~(\ref{ad3}) for slant depths $R>R_{12}$ one must substitute $R \to (R-R_{12})$ 
and $E_f \to E_{12}$ and use the values ($\alpha_2,\beta_2$) for a loss description. For slant depths $R \leq R_{12}$ the use of~
(\ref{ad3}) remains unchangeable and the loss values are expressed by ($\alpha_1,\beta_1$).
This algorithm may be extended to computations with any number of slopes of the energy losses.

The 5 parameters ($D_{0_\pi},D_{0_K},E^{cr}_{0_\pi}(0^\circ),E^{cr}_{0_K}(0^\circ),\gamma$) are those of the 
differential sea level muon spectrum, for which we use the following parametrization:
\begin{equation}\label{Sealds1}
D(E_{0},\theta)= E_{0}^{-\gamma} \sum_{i={\pi,K}} \frac{D_{0_i}}{1+E_{0}/E^{cr}_{0_i}(\theta)},  
\end{equation}
where $\gamma$ is a spectral index and $E^{cr}_{0_{\pi,K}}(\theta)$ have approximate sense of critical energies 
of pions and kaons for given zenith angle and $E^{cr}_{0_{\pi,K}}(0^\circ)$ are those for vertical direction.
The corresponding angular distrubution should be introduced using an analytical description of effective cosine $\cos\theta^*$  
taking into account the sphericity of atmosphere. It should be noted that the description of underwater angular flux with the
5 parameters of a sea level spectrum gives the possibility of their direct best fit by using the 
experimental underwater distribution.

Flux value in~(\ref{ad3}) is expressed in units of (cm$^{-2}$s$^{-1}$sr$^{-1}$) and all energies are in (GeV),
slant depth $R$ in units of (g$\,$cm$^{-2}$), loss terms $\alpha$ and $\beta$ 
in units of $(10^{-3}$GeVcm${}^2$g${}^{-1})$ and $(10^{-6}\mbox{cm}^2\mbox{g}^{-1})$, correspondingly.
\begin{figure}[t]
\vspace*{2.0mm} 
\includegraphics[width=8.3cm]{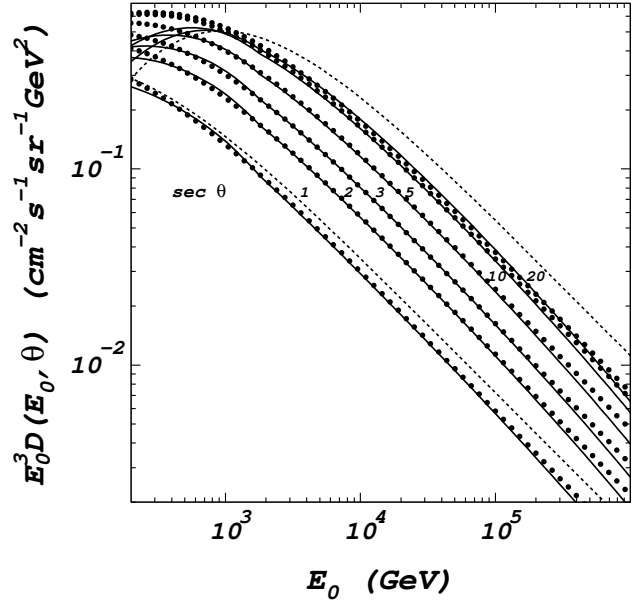} 
                                           
\protect\caption{
 Differential spectra of conventional muons at sea level for six zenith angles, sec$\,\theta$: 1, 2, 3, 5, 10, and 20, from bottom to top.  
 Curve labels correspond to values of sec$\,\theta$.  
 Solid curves: spectrum based on data tables from~\citep{Sineg}, dotted curves: spectrum defined by Eq.~(\protect\ref{BKfit}), 
 dashed curves: spectrum of Ref.~\protect\citep{VZK} shown only for two values of sec$\,\theta$: 1 and 20.  
\label{fig:Seal}}
\end{figure}
For the description of effective linear continuous energy losses we use the following values of parameters
when substituting in~(\ref{ad2}): ($\alpha_1$=2.67, $\beta_1$=3.40) and ($\alpha_2$=$-$6.5, $\beta_2$=3.66) with $E_{12}$=35.3 TeV.

To examine the angular behaviour of a flux given by the formula~(\ref{ad1}) by means of the comparison with numerical calculations 
we used the following parameters of the sea level muon spectrum: 
\begin{eqnarray*}
D_{0_\pi}=0.175,\qquad  D_{0_K}=6.475 \times 10^{-3},\\
E^{cr}_{0_{\pi}}(0^\circ)=103~\mbox{GeV},\quad E^{cr}_{0_{K}}(0^\circ)=810~\mbox{GeV}, \quad \gamma=2.72~.
\end{eqnarray*}
These values have been chosen according to splines computed in this work via the data tables kindly given us by 
authors of Ref.~\citep{Sineg}. 
When checking the values of fit spectrum for $\cos\theta$=(0.05--1.0) 
we realized that the standard description of effective cosine (with geometry of spherical
atmosphere and with definite value of effective height of muon generation) is not enough
and one should introduce an additional correction $S(\theta)$ leading to (10--20)$\,\%$
increase of effective cosine value for $\cos\theta<$~0.1. The reason of an appearing of this correction is that the concept of an
effective generation height is approximate one. It fails at large zenith angles where the real geometrical size of the generation region
becomes very large.

We should note that our expression~(\ref{Sealds1}) for the sea level muon spectrum does not contain a contrubution from
atmospheric prompt muons. According to the most recent calculations based on perturbative QCD, this contribution becomes 
essential only at $E_0 > 10^6$ GeV. The predictions of nonperturbative models    
are slightly more optimistic. We plan to generalize our approach and include this contribution in our following paper.
Incidentally, inclusion of prompt muons should be done in parallel with taking into account the steepening of the sea
level muon spectrum due to the knee in the primary cosmic ray spectrum.

The resulting fit of angular sea level spectrum in units of~(cm$^{-2}$s$^{-1}$sr$^{-1}$Ge$V^{-1}$) is given by
\begin{eqnarray}\label{BKfit}
D(E_0,\theta)=0.175 E_0^{-2.72} \nonumber \\
\times \left(\frac{1}{\displaystyle 1+\frac{E_0\cos\theta^{**}}{\displaystyle 103}}
+\frac{0.037}{\displaystyle 1+\frac{E_0\cos\theta^{**}}
{\displaystyle 810}}\right),
\end{eqnarray}
with modified effective cosine expressed by
\begin{equation}\label{Efcos}
\cos\theta^{**} =S(\theta)\cos\theta^{*},  
\end{equation}
where $\cos\theta^*$ is derived from spherical atmosphere geometry and is given by 
the polynomial fit:  
\begin{equation}\label{Efcosfit}
\cos\theta^{*}=\sum_{i=0}^4 c_{i}\cos^{i}\theta,  
\end{equation}
with the coefficients of the decomposition assembled in Table~\ref{tab:ectab}.
The accuracy of~(\ref{Efcosfit}) is much better than 0.3$\,\%$ except the region $\cos\theta$=(0.3--0.38) where
it may reach the value of 0.7$\,\%$.  
Note that for $\cos\theta>$~0.4 the influence of the curvature of real atmosphere is less
than 4~$\%$ but for $\cos\theta<$~0.1 it is greater than 40~$\%$. 

$S(\theta)$ is the correction which is given for $\sec\theta\leq20$ by
\begin{equation}\label{Scos}
S(\theta)=0.986+0.014\sec\theta.  
\end{equation}
Fig.~\ref{fig:Seal} illustrates the limits of applicability of angular spectrum given by Eq.~(\ref{BKfit}), for energy and zenith angle
variables.
The energy region, inside which the deviation from parent spectrum is less than 5~$\%$, is shifted from    
(0.3--200)$\,$TeV for $\cos\theta$=1.0 to (1.5--300)$\,$TeV for $\cos\theta$=0.05.  
The sea level spectrum given by~(\ref{BKfit}) is valid only below the knee ($E_{0}\sim\,$300 TeV) of primary cosmic ray spectrum.

Correspondingly, for critical energies in expression (3) one should use $\cos\theta^{**}$ instead of $\cos\theta^{*}$. 
\begin{table*}[htb]
\protect\caption{Coefficients $c_{i}$ of the fitting formula~(\protect\ref{Efcosfit})
                 for effective cosine with the maximum relative errors. }
\label{tab:ectab}
\center{\begin{tabular}{ccccccc} \hline \hline
$\cos \theta$ & $c_0$  & $c_1$ & $c_2$ & $c_3$ & $c_4$ & Max.err,$\%$ \\\hline
0$\div$0.002    & 0.11137 & 0 & 0 & 0 & 0 & 0.004\\ 
0.002$\div$0.2  & 0.11148 & $-0.03427$ & 5.2053  & $-14.197$ & 16.138 & 0.3\\ 
0.2$\div$0.8    & 0.06714 & 0.71578  & 0.42377 & $-0.19634$ & $-0.021145$ & 0.7\\\hline \hline 
\end{tabular}}
\end{table*} 
\section{Comparison with numerical calculations}
The examination of~(\ref{ad3}) showed rather quick convergence of series S$(z,\gamma)$ with increase of $R$ and $E_f$. 
Therefore, for the accuracy of $F_{cl}$ computation better than 0.1~$\%$ it is quite enough to take only 
four first terms of this series (up to $z^3$) for all values $R>$~1 km and $E_f$ in (1--10$^{4}$) GeV. Even using the two
terms leads to the accuracy of 1.3~$\%$ for ($R$=1.15 km, $E_f$=1 GeV) and $<$0.5$\,\%$ for ($R>\,$2.5 km, $E_f>\,$1 GeV).
 \begin{figure}[t]
 \vspace*{2.0mm} 
 \includegraphics[width=8.3cm]{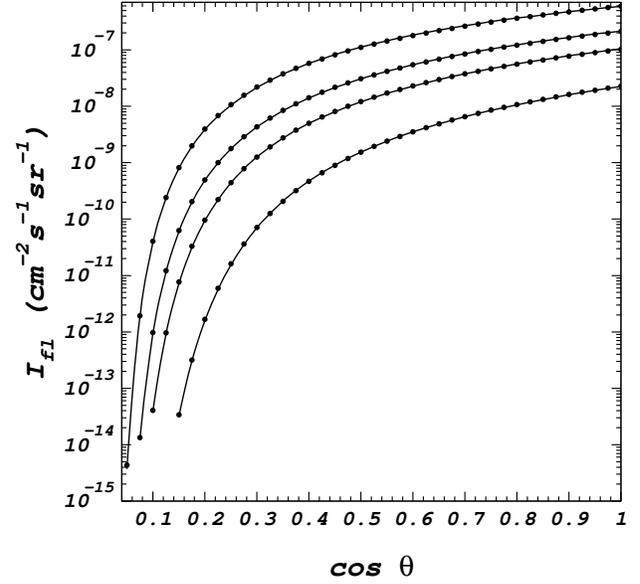}
 \protect\caption{
 Underwater integral muon flux allowing for loss fluctuations as a function of zenith angle at different vertical depths.
 for cut-off energy $E_f$=10 GeV. 
 Four curves correspond to vertical depths $h$: 1.15 km, 1.61 km, 2.0 km, and 3.0 km, from top to bottom. 
 Solid curves result from numerical computations by using the sea level spectrum based on data tables from~\citep{Sineg} 
 and MUM code of muon propagation.
 The dots are result of analytical calculation with expression~(\protect\ref{ad1}) by using the sea level spectrum~(\protect\ref{BKfit}). 
 \label{fig:form}}
 \end{figure}
\begin{figure}[t]
\vspace*{2.0mm} 
\includegraphics[width=8.3cm]{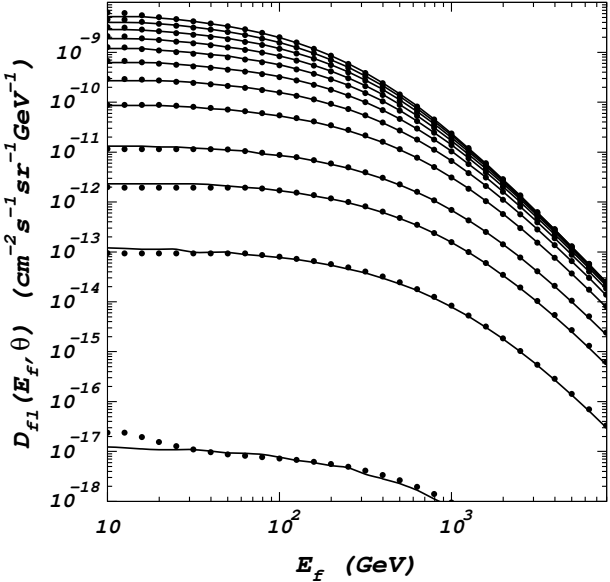} 
                                           
\protect\caption{
Underwater differential muon spectrum allowing for loss fluctuations as a function of energy $E_f$ at vertical depth of 1.15 km.
The curves are given for twelve zenith angles $\cos\theta$: 1.0, 0.9, 0.8, 0.7, 0.6, 0.5, 0.4, 0.3, 0.2, 0.15, 0.1, and 0.05, 
from top to bottom.  
Solid curves result from numerical differentiation of integral flux computed by using the sea level spectrum based on 
data tables from~\citep{Sineg} 
and MUM code of muon propagation.
Dotted curves result from numerical differentiation of the analytical expression~(\protect\ref{ad1}) 
based on the sea level spectrum~(\protect\ref{BKfit}).
\label{fig:diff_ef}}
\end{figure}
Fig.~\ref{fig:form} shows the comparison of underwater angular integral fluxes allowing for loss fluctuations 
at different basic depths $h$ (of location of existing and planned telescopes) calculated both numerically 
using MUM code~\citep{KBS2} for parent sea level spectrum  
and analytically~(\ref{ad1}) for the spectrum given by~(\ref{BKfit}). 

We realized that the error given by formula~(\ref{ad1}) for     
all mentioned sea level spectra is within the corridor of $\pm$(4--6)$\,\%$ for all cut-off 
energies $E_f$=(1--10$^{3})\,$GeV and
slant depths $R$=(1--16)$\,$km (corresponding angle is expressed by $\cos\theta=h/R$ for a given vertical depth $h$). This is proved
for $h$ in a range (1--3)$\,$km. 
For bigger cut-offs of $E_f$=(1--10)$\,$TeV the corridor of errors is $\pm$(5--7)$\,\%$ for $R$=(1--13)$\,$km. Note that for the
sea level spectrum~(\ref{BKfit}), just used for $C_f$ parametrization, the errors are smaller on 2$\,\%$. 

The accuracy of the parametrization, used for the correction factor as a function of $E_f$ and slant depth $R$   
is rather high and is about $\pm5\,\%$ for all angles and kinds of the sea level spectrum (assuming that the spectral
index $\gamma$ is approximately within (2.65--2.78)). It results in the possibility to use it for an estimating
numerically from various sea level spectra the value of an angular integral flux allowing for fluctuations of losses without
direct Monte Carlo simulations.

Note that the expression~(\ref{ad1}) may be directly used for an ice after substitution $R \to R/ \rho$, with $\rho$ 
being the ice density, and, with an additional error of $\sim2\,\%$, for a sea water. In spite of seeming complexity of the formulae~(\ref{ad1}),
~(\ref{ad2}) and~(\ref{ad3}) they may be easily programmed.

The validity of proposed formula up to cut-off energies 10 TeV allows a calculation of underwater angular differential
spectrum $D(E_{f},R,\theta)$ by means of numerical differentiation of expression~(\ref{ad1}). It leads to rather appropriate 
results up to slant depths (11--12)$\,$km. 
We illustrate this in Fig.~\ref{fig:diff_ef} by comparison the underwater angular spectra calculated by numerical 
differentiation of integral flux computed by using the sea level spectrum based on data tables from~\citep{Sineg} 
and MUM code of muon propagation, and numerical differentiation of the analytical expression~(\protect\ref{ad1}) 
based on the sea level spectrum~(\protect\ref{BKfit}).
\balance
For a vertical depth $h$=1.15 km it leads to errors
$\pm4\,\%$ for $E_f$=(20--$8\times10^{3})\,$GeV for the angles corresponding to slant depths $R=h/\cos\theta$ of (1--3)$\,$km
and $\pm$(6--8)$\,\%$ for $E_f$=(30--$5\times10^{3})\,$GeV for the slant depths (3--12)$\,$km. Even for $R$=23.2 km the result
is still valid within $\pm10\,\%$ but for the very narrow energy region $E_f$=(90--300)$\,$GeV. 
\section{Conclusions}
\label{sec:Concl}
The analytical expression presented in this work allows to estimate for fluctuating losses the integral flux of
atmospheric muons in pure water expected for different zenith angles, $\cos\theta$=(0.05--1.0), at various vertical
depths at least of $h$=(1--3)$\,$km for different parametrizations of the sea level muon spectra. 
The errors of this expression are estimated to
be smaller than $\pm$(4--6)$\,\%$ for cut-off energies ranged in $E_f$=(1--10$^{3})\,$GeV and slant depths 
in $h/\cos\theta$=(1--16)$\,$km. The main advantage of the presented formula consists in the possibility of the direct best fit of 5
parameters of parent sea level spectrum using angular distribution of underwater integral flux measured experimentally at a given
vertical depth.  

The validity of this analytical expression with an accuracy of $\pm$(5--7)$\,\%$ for $E_f$=(10$^3$--10$^4$)$\,$GeV 
and slant depths of (1--12)$\,$km gives also the possibility of estimation the angular underwater differential 
spectrum (by means of numerical differentiation) with error smaller than $\pm$(6--8)$\,\%$ for 
energies of (30--5$\times10^{3}$)$\,$GeV.

The accuracy of the parametrization, used for the correction factor as a function of $E_f$ and slant depth $R$   
is rather high and is about $\pm5\,\%$ for all angles and kinds of the sea level spectrum (assuming that the spectral
index $\gamma$ is approximately within (2.65--2.78)). It results in the possibility to use it for an estimating
numerically from various sea level spectra the value of an angular integral flux allowing for fluctuations of losses without
direct Monte Carlo simulations.

The proposed method may be adapted to estimations in rock after corresponding description of the correction factor and continuous effective losses.
\begin{acknowledgements}
We are grateful to V. A. Naumov for useful advices and to S. I. Sinegovsky and T. S. Sinegovskaya for making available
the muon sea level spectrum data tables.
\end{acknowledgements}


\begin{thebibliography}{99}
\bibitem[Klimushin, Bugaev, and Sokalski(2000)]{KBS} 
          S.~I.~Klimushin, E.~V.~Bugaev, and I.~A.~Sokalski, 
          hep-ph/0012032.
\bibitem[Klimushin, Bugaev, and Sokalski(2001)]{KBS1} 
          S.~I.~Klimushin, E.~V.~Bugaev, and I.~A.~Sokalski, 
          these Proceedings.
\bibitem[Misaki {\it et al.}(1999)]{Sineg} 
          A.~Misaki {\it et al.}, in {\em Proceedings of the 26th ICRC},
          Salt Lake City, Utah, 1999, edited by D.~Kieda, M.~Salamon, and B.~Dingus.
          Vol.~{\bf 2}, p.~139.
\bibitem[Sokalski, Bugaev, and Klimushin(2000)]{KBS2} 
          I.~A.~Sokalski, E.~V.~Bugaev, and S.~I.~Klimushin,  
          hep-ph/0010322.
\bibitem[Volkova, Zatsepin, and Kuz'michev(1979)]{VZK}
         L.~V.~Volkova, G.~T.~Zatsepin, and L.~A.~Kuz'michev, Yad.
         Fiz. {\bf 29}, 1252 (1979) [Sov. J. Nucl. Phys. {\bf 29},
         645 (1979)].
\end{thebibliography}
\end{document}